\documentclass[a4paper]{jpconf}
\usepackage{graphicx}

\begin{document}
\title{Leptons, quarks, and their antiparticles\\
from a phase-space perspective}

\author{Piotr \.Zenczykowski}

\address{Division of Theoretical Physics,
Institute of Nuclear Physics,
Polish Academy of Sciences,
Radzikowskiego 152,
31-342 Krak\'ow, Poland}
\ead{piotr.zenczykowski@ifj.edu.pl}

\begin{abstract}
It is argued that antiparticles may be interpreted in macroscopic terms
without explicitly using the concept of time and its reversal. 
The appropriate framework is that of
nonrelativistic phase space.
It is recalled that a quantum version of this approach
leads also, 
alongside the appearance of antiparticles,  to the 
emergence of `internal'  
quantum numbers identifiable with weak isospin, weak hypercharge and colour,
and to the derivation of the Gell-Mann-Nishijima relation, while
simultaneously  offering
a preonless interpretation of the
Harari-Shupe rishon model. Furthermore, it is shown that -
under the assumption of the additivity of canonical momenta -
the approach entails the emergence of 
string-like structures resembling mesons and baryons, thus
providing a different starting point
for the discussion of quark unobservability.
 \end{abstract}

\begin{flushright}
{\parbox{270pt}{\small {{\it `It is utterly beyond our power to measure the changes of things by time
(...) time is an abstraction at which we arrive by means of the changes of
things; made because we are not restricted to any one definite measure, all
being interconnected.'}
\begin{flushright}Ernst Mach - \cite{Mach}\end{flushright}}}}
\end{flushright}

\section{Introduction}
\subsection{Charge conjugation and time}
In the Standard Model (SM), elementary particles
are grouped into
multiplets of various symmetry groups such as $SU(2)$, $SU(3)$, etc.
Particles and antiparticles belong then to
complex conjugate representations (e.g. when coloured quarks are assigned to
representation $\bf 3$ of $SU(3)_C$, the antiquarks are assigned to $\bf 3^*$). 
With the standard particle theory formulated on the background of 
classical space and time, the concepts of
 complex conjugation and time reversal are closely related.
 Accordingly, St\"uckelberg and Feynman interpret the antiparticles
  as particles `moving backwards in time'.
  
  The spacetime-based description of Reality provided by the Standard Model
 is very successful.
Yet, there are many questions that go beyond what SM was
devised to answer. These include: why the physical quantities that enter 
into the SM as parameters
have their specific values (masses of elementary particles or mixing angles
between particle generations), 
what is the origin of various internal quantum numbers,
or how to describe both elementary particles and gravity in
a single framework.

The latter issue suggests that the standard particle theory should give way to an
approach in which spacetime serves no longer as a background but 
becomes a dynamical structure,
in line with general relativity ideas.
In fact, many physicists have argued 
that we need an approach in which macroscopic
classical
time and space emerge 
- in the limit of a large interconnected structure -
 out of a simple timeless
and alocal quantum level \cite{pregeometry},\cite{Finkelstein}. 
The description of this level is still expected to be complex,
 as quantum descriptions are supposed to be \cite{BarbourComplexQG}.
  
Can one interpret then classically the expected
presence of particles and antiparticles 
without explicitly using the concept of
macroscopic time? 
In order to deal with this question, 
we find it appropriate to recall first how the classical
concept of background time  
was originally introduced.

\subsection{Time and change}
In classical approaches with background time,  
position is viewed as a function
of time ${\bf x}(t)$, while its change $\delta {\bf x}$
is thought of as occurring in time, {\it `due'} to its increase: 
$\delta {\bf x} = \frac{d{\bf x}}{dt}\delta t$. However, a
deeper insight undermines this concept of background time.
Rather, it views time as an effective parameter that somehow parametrizes
{\it change}, as stated by Ernst Mach (see motto).

In \cite{BarbourDICE2002andFQXi} Barbour described how energy
conservation (in an isolated system) serves as a key ingredient 
that leads to a definition of an increment
of ephemeris time through observed changes:
\begin{equation}
\label{encons}
\delta t =\sqrt{\frac{\sum_im_i(\delta {\bf x}_i)^2}{2(E-V)}},
\end{equation}
where $\delta {\bf x}_i$ are measured changes in the positions of astronomical 
bodies, $E$ is total (and fixed) energy  and $V$ is the gravitational potential
of all interacting bodies of the system.
Thus,  the (astronomical) time is defined 
by change, not vice versa.
A more neutral expression of the relation between 
an increment of time $\delta t$ 
and changes of position  $\delta {\bf x}_i$ is that the
two are correlated. It is then up to us to decide which of the two alternative
formulations to choose: change in (background) time, or time 
defined from (observed)
 change.\\

Now, in full analogy with Eq. (\ref{encons}), it is natural to consider
\begin{equation}
\delta t ~{\bf P} = \sum_i m_i ~\delta {\bf x}_i,
\end{equation}
where ${\bf P}$ is the total (and conserved) momentum of the system.
For the sake of our discussion, we 
restrict the above equation to the one-particle case
\begin{equation}
\label{tdef}
{\bf p}~ \delta t = m ~\delta {\bf x},
\end{equation}
and observe that, again, one may view this relation in two ways:\\ 
1) either with momentum ${\bf p}$ calculated 
from the change of position $\delta {\bf x}$
in a given increment of (background) time $\delta t$, or \\
2) with time increment $\delta t$ calculated from given 
${\bf p}$ and $\delta {\bf x}$.\\

The latter standpoint - in which momentum is not calculated 
from a change in position but assumed as independent of position - 
is familiar from the Hamiltonian formalism, in which
momenta and positions are treated as independent variables.
Thus, the idea of time induced by change should be
expressible in the language of
 phase space in which
the macroscopic arena is not 3- but 6- dimensional. 
Furthermore, with independent ${\bf p}$ and ${\bf x}$ one may 
consider various independent transformations of ${\bf p}$ and ${\bf x}$, 
for example, study 
the symmetries of the time-defining equation Eq.(\ref{tdef}).
Then, standard 3D reflection corresponds to $({\bf p},{\bf x}) \to (-{\bf
p},-{\bf x})$ 
(and leaves time untouched),
while the operation $({\bf p},{\bf x}) \to ({\bf p},-{\bf x})$ 
leads to time reflection: $t \to -t$.\\

Moving now to the quantum description we observe that
 the canonical commutation relations
 naturally involve the imaginary unit (we use units in which $\hbar =1$):
 \begin{equation}
 \label{eq1}
 [x_k,p_l]=i \delta_{kl}.
 \end{equation}
In spite of containing $i$, 
Eq. (\ref{eq1}) does not involve time explicitly. 
Consequently, since charge and complex conjugations are related,  
it should be possible to give
an interpretation to the particle-antiparticle degree of freedom
using the phase-space concepts of positions and momenta alone, i.e.
 without referring to the concept of explicit time.
Charge conjugation should then be seen 
not only as connected to time reversal,
but also
as just one of several possible 
transformations in phase space.

\section{Noncommuting phase space}
In the standard picture
with background spacetime there is a connection
between the properties of the background (e.g. under 3D rotation) 
and the existence of the 
`spatial' quantum numbers (e.g. spin).
Therefore, if one is willing to enlarge the arena to that of phase space, 
with
independent position and momentum coordinates, one might expect the appearance
of additional quantum numbers \cite{PLB2008}. 
From the point of view of the standard (3D space + time)  
formalism, such quantum numbers would necessarily appear 
`internal'.
\subsection{Born reciprocity}
The issue of a possible relation between particle properties (such as 
quantum numbers or masses), and the concept of  phase space was of high concern 
already to Max Born. 
In his 1949 paper \cite{Born1949} he discussed the difference between 
the concepts of position and momentum for elementary particles and
noted that the notion of mass appears in the relation
$p^2=m^2$, while $x^2$, the corresponding invariant in coordinate space 
(with $x^2$ of atomic dimensions),
does not seem to enter in a similar relation.
At the same time, Born stressed that various laws of nature such as
\begin{eqnarray}
&&\dot{x}_k=\frac{\partial H}{\partial p_k},~~~
\dot{p}_k=-\frac{\partial H}{\partial x_k},\nonumber\\
&&{}[x_k,p_l]=i\delta_{kl},\nonumber\\
&&{}L_{kl}=x_kp_l-x_lp_k,
\end{eqnarray}
are invariant under the `reciprocity' transformation:
\begin{eqnarray}
x_k \to p_k,&&p_k \to -x_k.
\end{eqnarray}
 Noting that relation
 $p^2=m^2$ is not invariant under this transformation, he concluded: 
 {\it `This lack of symmetry seems to me very strange and
 rather improbable'}.

The simplest phase-space
generalization of the 3D (nonrelativistic) concepts of rotation and reflection
is obtained with a symmetric treatment of the two O(3) invariants,
${\bf x}^2$ and ${\bf p}^2$, via
 their addition
(obviously, this procedure
requires an introduction of a new fundamental
constant of Nature of dimension [momentum/distance]):
\begin{eqnarray}
\label{basicinv}
&{\bf x}^2+{\bf p}^2.&
\end{eqnarray}
Eq.(\ref{basicinv}) is invariant under $O(6)$ transformations 
and Born reciprocity in particular. 

We may now treat  ${\bf x}$ and ${\bf p}$ as operators, and
require their commutators to be form invariant. 
The original $O(6)$ symmetry 
is then reduced to $U(1)\otimes SU(3)$. 
The appearance of the symmetry group present 
in the Standard Model
leads us to ask the question whether phase-space symmetries could possibly lie at the
roots of SM symmetries.
As shown in \cite{PLB2008} and also argued below, this seems quite
possible.

\subsection{Dirac linearization}
Our present theories are standardly divided into classical and quantum
ones. Yet, as stressed by Finkelstein, the latter should be more appropriately
regarded as belonging to a mixed, classical-quantum type. 
In Finkelstein's view, the
classical (c), and classical-quantum (cq) theories should give way to a purely
quantum (q) approach in which infinities would not appear and 
the concept of infinite and infinitely divisible background space would no longer
exist \cite{Finkelstein}. This view is clearly in line with the arguments 
 for
 a background-independent approach to quantum gravity.

The Dirac linearization prescription may be thought of as a procedure 
that has led us from the 
classical description of Nature to part of its quantum description.
Indeed, the linearization of ${\bf p}^2$ leads to the appearance of Pauli 
matrices, which describe spin at the quantum level.
It is therefore of great interest to apply Dirac's idea to the phase-space
invariant of Eq.(\ref{basicinv}).
Using anticommuting matrices 
$A_k$ and $B_k$ ($k=1,2,3$) (for more details, see
\cite{PLB2008})
one finds
\begin{equation}
({\bf A}\cdot {\bf p}+{\bf B}\cdot {\bf x})
({\bf A}\cdot {\bf p}+{\bf B}\cdot {\bf x})=
({\bf p}^2+{\bf x}^2)+ \sum_1^3 \sigma_k\otimes\sigma_k\otimes\sigma_3
\equiv R+R^{\sigma}.
\end{equation}
The first term on the r.h.s., $R={\bf p}^2+{\bf x}^2$, 
appears thanks to the anticommutation properties of $A_k$ and $B_l$. 
The other term, $R^{\sigma}$, appears because
$x_k$ and $p_k$ do not commute.
These two terms sum up to a total $R^{tot}=R+R^{\sigma}$.

When viewed from Finkelstein's perspective, the invariant
\begin{equation}
\label{qandcqconnection}
{\bf A}\cdot {\bf p}+{\bf B}\cdot {\bf x}
\end{equation}
connects then the cq-level of phase space 
(noncommuting positions ${\bf x}$ and momenta ${\bf p}$)
with the presumably purely q-level
structure: the Clifford algebra built from matrices ${\bf A}$ and ${\bf B}$.

\section{Quantization}
\subsection{Gell-Mann-Nishijima relation}
Just as $R$ is quantized, so is $R^{\sigma}$. Thus, we have to
find the eigenvalues of $R^{\sigma}$.
For better correspondence with the standard definitions 
of internal quantum numbers,
we introduce operator $Y$:
\begin{equation}
Y\equiv \frac{1}{3}R^{\sigma}B_7=\frac{1}{3}\sum_1^3
\sigma_k\otimes\sigma_k\otimes\sigma_0\equiv 
\sum_1^3Y_k.
\end{equation}
where $B_7=iA_1A_2A_3B_1B_2B_3$ is 
the 7-th anticommuting element of the Clifford algebra.

\begin{table}
\caption{Decomposition of eigenvalues of Y into eigenvalues of its
components.}
\label{table2}
\begin{center}
\begin{math}
\begin{array}{lllll}
\br
{\rm colour~~}  & ~~0 & ~~1 & ~~2 & ~~3 \\
\mr 
Y &-1&+\frac{1}{3}&+\frac{1}{3}&+\frac{1}{3}
\rule{0mm}{5mm}\\
Y_1 &-\frac{1}{3}&-\frac{1}{3}&+\frac{1}{3}&+\frac{1}{3}
\rule{0mm}{5mm}\\
Y_2 &-\frac{1}{3}&+\frac{1}{3}&-\frac{1}{3}&+\frac{1}{3}
\rule{0mm}{5mm}\\
Y_3 &-\frac{1}{3}&+\frac{1}{3}&+\frac{1}{3}&-\frac{1}{3}
\rule{0mm}{5mm}\\
\br
\end{array}
\end{math}
\end{center}
\end{table}
Since $Y_k$ commute among themselves, they may be
simultaneously diagonalized. The 
eigenvalues of $Y_k$ ($k=1,2,3$) are $\pm 1/3$. The
resulting pattern of possible eigenvalues of $Y$ is shown in Table
\ref{table2}.

In \cite{PLB2008} a conjecture was put forward that the electric charge 
$Q$ is proportional to operator $R^{tot}B_7$, evaluated
for the lowest level of $R$, i.e.:
\begin{equation}
\label{GMN}
Q=\frac{1}{6}(R_{lowest}+R^{\sigma})B_7=I_3+\frac{Y}{2},
\end{equation}
where $R_{lowest}=({\bf p}^2+{\bf x}^2)_{lowest}=3$, and
$I_3=B_7/2$.

The above equation 
 is known under the name of the Gell-Mann-Nishijima relation 
 (with $I_3$ of eigenvalues $\pm 1/2$ known as weak isospin and Y of eigenvalues
 $-1,+1/3$ known as weak hypercharge)
 and is considered to be a law of nature. 
It summarizes the pattern of charges of
 all eight leptons and quarks from a single
SM generation.
In the phase-space approach it is {\it derived} as a consequence of phase-space
symmetries.

\subsection{Harari-Shupe rishons}
As shown in \cite{PLB2008}, the pattern
in which the weak hypercharge $Y$ is built out of `partial hypercharges'
$Y_k$ corresponds {\it exactly} to the pattern in which electric charges are
built in the Harari-Shupe (HS) model of quarks and leptons \cite{Harari}.
The HS approach describes the structure of a SM
generation with the help of a composite model:
it builds all eight fermions of a single generation from 
two spin-$1/2$ `rishons' $V$ and $T$ of charges
0 and +1/3. 
The proposed structure of leptons and quarks is
 shown in Table \ref{table1}.

\begin{table}
\caption{Rishon structure of leptons and quarks with $I_3=+1/2$.}
\label{table1}
\begin{center}
\begin{math}
\begin{array}{lllllllll}
\br
&\nu_e~~~&u_R~~&u_G~~&u_B~~&e^+~~&
\bar{d}_R~~&\bar{d}_G~~&\bar{d}_B~~\\
\mr 
&VVV&VTT&TVT&TTV&TTT&TVV&VTV&VVT\rule{0mm}{4mm}\\
Q~~ &~~0&+\frac{2}{3}&+\frac{2}{3}&+\frac{2}{3}&+1
&+\frac{1}{3}&+\frac{1}{3}&+\frac{1}{3}
\rule{0mm}{4mm}\\
Y~~ &-1&+\frac{1}{3}&+\frac{1}{3}&+\frac{1}{3}&+1
&-\frac{1}{3}&-\frac{1}{3}&-\frac{1}{3}
\rule{0mm}{4mm}\\
\br
\end{array}
\end{math}
\end{center}
\end{table}

Our phase-space approach not only reproduces exactly the successful part of the
rishon structure, but it also removes all the main shortcomings 
of the HS model.
In particular, the approach is {\it preonless}, i.e. 
the phase-space `rishons' are components of charge (hypercharge) only, 
with {\it no
interpretation in terms of spin 1/2 subparticles}. Thus,
there is no problem of rishon confinement.
Consequently, our leptons and quarks are viewed as pointlike, 
in perfect agreement with the experimental knowledge.
In summary, the phase-space approach explains the origin of the observed symmetries 
without introducing any subparticles.

One might object 
that the nontrivial combination of spatial and internal symmetries 
 is forbidden by the Coleman-Mandula no-go theorem \cite{CMnogo}. 
 Yet, this theorem is neatly evaded by our construction:
 the theorem works at the S-matix level, while quarks are to be
 confined, as we expect and as will be argued later on.
 Furthermore, no additional 
 dimensions (standardly understood) have actually been added in our framework. 
 The only 
 change was a shift in the conceptual point of view: 
 instead of a picture based on 3D space
 and time, we decided to view the world in terms of a picture based on the 6D
 arena of canonically conjugated positions and momenta.
 
\section{Compositeness and additivity}
The linearized phase-space approach suggests that the Clifford algebra of
nonrelativistic phase space occupies an important place in our description
of leptons and quarks (see also \cite{Clifford}). Thus, one would like to use it also in a
description of composite systems, 
in particular in the description of hadrons as composite systems of
quarks. Unfortunately, it is not clear how to achieve this goal.
Yet, the q-level construction in question
 must be related
to the c-level description of Nature, 
and, consequently, should be interpretable at the purely classical level. 
Indeed, as Niels Bohr said: {\it `However far the phenomena transcend the scope of
classical physical explanation, the account of all evidence must be expressed in
classical terms'}. In fact,
Eq.(\ref{qandcqconnection}) provides the required connection through which
 transformations between leptons, quarks, and their antiparticles
are related to those in phase space.

Now, in any description of composite systems, be it at the classical or the quantum
level,
an important ingredient 
is provided by the tacitly assumed concept of additivity. 
Indeed, additivity is assumed both
at the quark level (e.g. additivity of spins or flavour quantum
numbers),
and at the classical level (e.g. additivity of momenta).
The question then appears: can this property of the additivity of momenta
at the classical level
be somehow used to infer about the properties of such 
q-level objects as our quarks when these are viewed from the
macroscopic classical perspective.

The relevant basic macroscopic observation  is that for any direction
in 3D space, and completely regardless of what happens in the remaining two
directions, we have the
{\it additivity of physical momenta of any number of ordinary particles or
antiparticles, quite irrespectively of their internal quantum numbers}
(e.g. $P_z=\sum_ip_{iz}$).
(By ordinary particles we mean those that can be oberved 
{\it individually}, such as
leptons and hadrons, but not quarks.)
Note that the positions of ordinary particles 
are not additive in such a simple way, since a composite
object is best described in terms of its center-of-mass coordinates:
$X_z =\sum_i m_ix_{iz}/\sum_i m_i$.
In the following, we shall present the connections between lepton-quark and
phase-space transformations and discuss their implications for 
the concept of the additivity of momenta.

\subsection{Charge conjugation}
In a quantum description the transition from particles to antiparticles 
is effected by complex
conjugation. 
Consider now a system of particles in which some
 are being transformed into antiparticles.
We want to find the related transformation in phase space. 
We note that in order to preserve the principle of the
additivity of
physical momenta, one is not allowed to change
the momenta of any of the particles being transformed.
The invariance of $[x_k,p_l]=i\delta _{kl}$ requires then
that
\begin{eqnarray}
p_k \to p_k,~~ &~~ x_k \to -x_k,~~ &~~ i \to -i.
\end{eqnarray}
Using the invariance of Eq.(\ref{qandcqconnection}), the ensuing
transformations of $A_k$ and $B_k$, 
and the definition of $Q$, $I_3$, and $Y$, 
one can check \cite{Zen2010} that in this way we are indeed led from
particles to antiparticles. Thus, in the phase-space picture, with time regarded as
secondary, the antiparticles are related to particles via $i \to -i$
combined with
the reflection of position space (with 
the time reflection $t \to -t$
induced via the invariance of Eq.(\ref{tdef})).

\subsection{Isospin reversal}
Similarly, one checks \cite{Zen2010} that isospin reversal $I_3 \to -I_3$ corresponds
to the transformation $A_k \to A_k, B_k \to -B_k$, and $i \to i$, and,
consequently, to $p_k \to p_k$ and $x_k \to -x_k$. Yet,
this is not the same as charge conjugation since in this case, with $i \to +i$, the
momentum-position commutation relations do not stay invariant. 

In summary, the phase-space 
representations of both the particle-antiparticle transformation and of the
isospin reversal {may be} (and have been) chosen in a way that 
does not affect the momenta of any of the particles, thus 
preserving their additivity.

\subsection{Lepton-to-quark transformations}
Transformations from the lepton to the quark sector 
(in particular, the change
$Y=-1 \to Y=+1/3$) require the use of 
SO(6) rotations going outside those generated by
the familiar $1+8$ generators of $U(1) \times SU(3)$.
The remaining six of the fifteen SO(6) generators form two SU(3) triplets, of
which only one actually leads to transformations of some $A_k$ into $B_m$, 
while keeping $i$ and $I_3$ fixed \cite{PLB2008}. 

Under these transformations, ${\bf A}$ is changed into some 
${\bf A}^{Qn}$, and ${\bf B}$ into some ${\bf B}^{Qn}$, with
$n=1,2,3$ being the colour index \cite{Zen2010}.
For the transformed elements 
${\bf A}^{Qn}$ (and similarly for ${\bf B}^{Qn}$)
one can then choose the following (not unique) 
representation :
\begin{eqnarray}
\label{quarkrepresentation}
A^Q\equiv 
\left[
\begin{array}{c}
{\bf A}^{Q1}\\
{\bf A}^{Q2}\\
{\bf A}^{Q3}
\end{array}
\right]
&=&
\left[
\begin{array}{ccc}
A_1&-B_3&+B_2\\
+B_3&A_2&-B_1\\
-B_2&+B_1&A_3
\end{array}
\right],
\end{eqnarray}
with the 
quark {\it canonical momenta} ${\bf P}^{Qn}$ (and likewise, for 
quark canonical positions ${\bf X}^{Qn}$), 
obtained from the condition of the invariance of Eq.(\ref{qandcqconnection}):
\begin{eqnarray}
\label{canmomrepresentation}
P^Q\equiv 
\left[
\begin{array}{ccc}
{{\bf p}}^{Q1}\\
{\bf p}^{Q2}\\
{\bf p}^{Q3}
\end{array}
\right]
&=&
\left[
\begin{array}{ccc}
p^1_{1}&-x^1_{3}&+x^1_{2}\\
+x^2_{3}&p^2_{2}&-x^2_{1}\\
-x^3_{2}&+x^3_{1}&p^3_{3}
\end{array}
\right].
\end{eqnarray}
The above forms of $A^Q$ and $P^Q$ remain unchanged
for quarks of opposite isospin.

For the antiquarks, the relevant forms are
 \begin{eqnarray}
A^{\overline{Q}}
&=&
\left[
\begin{array}{ccc}
A_1&+B_3&-B_2\\
-B_3&A_2&+B_1\\
+B_2&-B_1&A_3
\end{array}
\right],
\end{eqnarray}
and
 \begin{eqnarray}
 \label{canmomantiquarks}
P^{\overline{Q}}
&=&
\left[
\begin{array}{ccc}
{p}^1_{1}&+{x}^1_{3}&-{x}^1_{2}\\
-{x}^2_{3}&{p}^2_{2}&+{x}^2_{1}\\
+{x}^3_{2}&-{x}^3_{1}&{p}^3_{3}
\end{array}
\right].
\end{eqnarray}
Again, these forms are independent of isospin.

As expected, 
the difference between quarks and antiquarks is represented
by a change in sign in front of physical positions entering into the
 definitions of quark (antiquark) canonical momenta.
{\it If} additivity of canonical momenta of quarks is a proper generalization 
 of the additivity
 of physical momenta for leptons and other individually observable
  particles, then - on account of the relative (positive and negative) signs between
 position components in (\ref{canmomrepresentation}) and 
 (\ref{canmomantiquarks}) - this additivity 
 (separately in each of the relevant phase-space directions) leads to
 translationally invariant string-like expressions for quark-antiquark and three-quark
 systems 
 (i.e. $x^1_2(q)-x^1_2(\overline{q})$ for $q\bar{q}$
 and  $(x^3_1(q_3)-x^2_1(q_2),x^1_2(q_1)-x^3_2(q_3),x^2_3(q_2)-x^1_3(q_1))$
 for $q_1q_2q_3$), but not for $qq$ or $qqqq$ systems.
 Additivity of canonical momenta leads therefore to the
  formation of `mesons' and `baryons' only.

 In principle, the chain of arguments leading to Eq.
 (\ref{quarkrepresentation},\ref{canmomrepresentation}) could involve ordinary
 reflections, e.g $(A_1,-B_3,+B_2) \to (A_1,+B_3,+B_2)$, {\it before} putting the
 latter expression
 and its cyclic counterparts together into a matrix form similar to
 (\ref{quarkrepresentation}) (`grouping'). 
 The corresponding phase-space counterparts would then look
 somewhat different, i.e.
 $(p_1,-x_3,+x_2) \to (p_1,+x_3,+x_2)$ etc. The translational
 invariance of three-quark systems could then not be achieved by simply adding
 the appropriate canonical momenta of different quarks, because all physical
 position coordinates would enter an analog of (\ref{canmomrepresentation}) with positive signs. 
 The point, however, is that 
 one {\it may} choose 
 the ordering of the operations of grouping and reflection
  in such a way that - by the simple
  procedure of addition -
  the translational invariance can be achieved at all.
  Furthermore, it seems nontrivial that this requires
  collaboration of phase-space representatives of quarks
  of three different colours. After all, the latter were
  originally defined - {\it in a way seemingly independent of the concept of
  additivity} - 
  via a diagonalization procedure performed at the Clifford
  algebra level.

 In other words, the q-level structure of coloured quark charges corresponds
 to a specific picture in the macroscopic arena.
 According to this picture, individual quarks are not observable at the
 classical level since their individual canonical momenta are not
 translationally invariant. On the other hand, 
 translational invariance may be restored via the
 collaboration of quarks of different colours. 
 When viewed from our classical
  point of view, the resulting composite 
 systems possess standard particle-like properties, while at the
 same time exhibiting internal string-like features. 
 Thus, quark unobservability is supposed to be
 connected to the very 
 emergence and nature of space and time.
 Speaking more precisely, quarks are supposed to be unobservable
 because space and phase space are most probably just convenient 
 classical abstractions, 
 into the descriptive corset of which we try to force various pieces of Reality.

The often-used argument that `space is standard' at the distances a few orders
of magnitude smaller than proton's size is not sufficiently sound.
After all, the existence of long-distance
nonlocal quantum correlations indicates that - at least for some purposes -
our classical spacetime
concepts (into which we try to force our descriptions of elementary 
particles) are 
inadequate 
at much larger scales. In fact, statements about an `unchanged
nature of space' at a distance
of $10^{-18}~m$ or so
 follow from the success of the Standard Model (a cq-level theory)
 and are strictly valid only {\it within} 
 the description it provides, not outside of it 
 (e.g. not in a q-level theory in
 which space is to be an emergent concept only).

The above discussion suggests that the phase-space approach has the capability
of describing the phenomenon of quark unobservability in a way 
seemingly different from the SM flux-tube picture of confinement. 
In fact, however, the phase-space approach does not have to be in conflict with the
latter picture, just as the Faraday picture of `real' fundamental field lines 
is not in conflict with the Maxwell concept of fields. 
Rather, we regard the idea of the linearized phase space as offering
a possible q-level starting point.   
Our discussion is based on the invariance of Eq. (\ref{qandcqconnection}) 
which provides a link between q- and cq- (c- ) levels of description.
Obviously, however, Eq. (\ref{qandcqconnection}) does not specify how the macroscopic
background phase space actually emerges from the underlying quantum level.
Hence, at present there is no way to compare our ideas directly
with the standard, background-dependent,
QCD-based picture of confinement. 

\section{Conclusions}
The linearized phase-space approach differs markedly from the 
standard frameworks, in
particular from the $SU(5)$-based unifications. 
A brief comparison of some differences between the two
schemes is given in Table \ref{table3}. 
\begin{table}[t]
\caption{Comparison with $SU(5)$.
}
\label{table3}
\begin{center}
\begin{math}
\begin{array}{ll}
\br
{{\rm {Simple~ group}}~-~ SU(5)}&{\rm Linearized~phase~space}\rule{0mm}{5mm}\\
\mr
5^*+10~(+1) ~{\rm representations} & 
{\rm two~replicas~of}~4~{\rm and}~4^*~{\rm of~} SU(4) \rule{0mm}{4.5mm}\\
{\rm fair~prediction~for~Weinberg~angle}& {\rm -} \rule{0mm}{4.5mm}\\
{\rm anomaly~free} & {\rm -}\rule{0mm}{4.5mm}\\
{\rm no~connection~to~macroscopic~arena~~~~~~~~}& {\rm connection ~to~
macroscopic~phase~space}\rule{0mm}{5mm}\\
{\rm proton~decay}&{\rm protons~are~forever}\rule{0mm}{4.5mm}\\
{\rm -}&{\rm Harari-Shupe~reproduced}\rule{0mm}{4.5mm}\\
\br
\end{array}
\end{math}
\end{center}
\end{table}
In the author's opinion, the SU(5) approach lacks a solid
philosophical background. On the other hand, the phase-space approach, although
overly simplistic, 
satisfies an important philosophical condition:
the necessity to connect the q-level description
of elementary particles to the c-level description of the
macroscopic world.
One can hear the echoes of this condition in the words of Roger Penrose,
 who stated in \cite{Penrose}:
{\it `I do not believe that a real understanding of the nature of
elementary particles can ever be achieved without a simultaneous deeper
understanding of the nature of spacetime.'}

The phase-space approach provides a possible theoretical explanation of
the structure of a single generation of the Standard Model. 
It gives us a tentative (pre)ge\-ometric
interpretation of the origin of the Gell-Mann-Nishijima
relation. 
It reproduces the structure of the Harari-Shupe preon model without actually
introducing any preons at all, in line with the standard pointlike description of
fundamental fermions.
It touches on the question of the origin of time.
It suggests that the phenomenon of quark unobservability is related
to the very nature of space and time, which are regarded as emergent
(or abstract) classical concepts only.   

The picture offered by the Clifford algebra of nonrelativistic phase space
 need not be regarded as
 `the' q-level approach.
Rather, it should be thought of as  
a possible deeper layer of description only.
However, it has encouraging features, which - as I believe -
will show up at the classical level if derived `in an emergent way' 
from any 
suitable 
q-level description.
Let me therefore end by paraphrasing Penrose's opinion: 
I do not believe that deeper understanding of elementary
particles can be achieved without further studies of the proposed link
between the elementary particles themselves and the properties and symmetries
 of nonrelativistic phase space.\\

This work has been partially supported by the Polish Ministry of Science and
Higher Education research project No N N202 248135.

\end{document}